\documentclass[runningheads]{llncs}
\usepackage{microtype}
\usepackage{graphicx}
\usepackage{xcolor}
\usepackage{caption}
\usepackage{gradientframe}
\usepackage[T1]{fontenc}
\usepackage[utf8]{inputenc}
\def\doi#1{\href{https://doi.org/\detokenize{#1}}{\url{https://doi.org/\detokenize{#1}}}}
\usepackage{listings}
\usepackage{amsmath}
\usepackage{amsfonts} 
\usepackage{pdfpages}
\usepackage{placeins}
\usepackage{pgfplots}
\usepackage{caption}
\usepackage{subcaption}
\usepackage[linesnumbered]{algorithm2e}
\usepackage{standalone}
\lstdefinestyle{lststyle}{%
    escapechar=@,
    columns=fullflexible
}
\lstdefinelanguage{complexity_lan}{
    keywords = {while, do, if, then, end,}
}
\definecolor{myGrey}{RGB}{140,140,140}
\lstdefinelanguage{pseudo_lan}{
    keywords = {while, do, if, then, end, for, else, return},
    comment=[l]{\#},
    commentstyle=\color{myGrey}\ttfamily
}
\begin{document}
\title{Accelerating Brain Simulations with the Fast Multipole Method}
%
%
\author{Hannah Nöttgen\inst{1,2}\orcidID{0000-0001-6481-5717} \and
Fabian Czappa\inst{1,3}\orcidID{0000-0001-8422-5706} \and
Felix Wolf\inst{1,4}\orcidID{0000-0001-6595-3599}}
\authorrunning{H. Nöttgen et al.}
%
\institute{Laboratory for Parallel Programming, Technical University of Darmstadt, Darmstadt, Germany
\and
\email{hannah.noettgen@stud.tu-darmstadt.de}
\and
\email{fabian.czappa@tu-darmstadt.de}
\and
\email{felix.wolf@tu-darmstadt.de}}
\maketitle              
\begin{abstract}
The brain is probably the most complex organ in the human body. 
To understand processes such as learning or healing after brain lesions, we need suitable tools for brain simulations. 
The Model of Structural Plasticity offers a solution to that problem. 
It provides a way to model the brain bottom-up by specifying the behavior of the neurons and using structural plasticity to form the synapses. 
However, its original formulation involves a pairwise evaluation of attraction kernels, which drastically limits scalability. 
While this complexity has recently been decreased to $O(n \cdot \log^2 n)$ after reformulating the task as a variant of an n-body problem and solving it using an adapted version of the Barnes--Hut approximation, we propose an even faster approximation based on the fast multipole method (FMM). 
The fast multipole method was initially introduced to solve pairwise interactions in linear time. Our adaptation achieves this time complexity, and it is also faster in practice than the previous approximation. 
\end{abstract}

\section{Introduction}
\label{sec:intro}
The human brain undergoes constant change not only in children but throughout the whole life~\cite{butz2009}.
These changes, especially in the form of synapse creation and deletion, are believed to be responsible for a major portion of the brain dynamics.
There is overwhelming evidence that structural plasticity, i.e., the change of connectivity of neurons, is responsible for learning, memory creation, and healing after lesions~\cite{diazpier2016,Keck2008,Holtmaat2009,Kleim13228,Kleim628}.   
However, current in vivo imaging techniques cannot create connectivity maps for human brains at a scale comparable to the original~\cite{beaujoin2018post,diazpier2016,dodt2007ultramicroscopy}.
This leaves a large portion of current research in need of simulations to fill the gap.
Many state-of-the-art simulators can mimic very complex behaviors of a single neuron, however, they lack the possibility to let neurons freely connect to others.
This task inherently involves solving pairwise interactions. Seeing that the human brain contains 86 billion neurons~\cite{Berlinn.d.}, this drastically limits scalability.
Many simulators bypass this issue by only allowing already existing connections to be strengthened or weakened (synaptic plasticity), bringing the complexity down to linear in the number of neurons and synapses per neuron.

It is not fully understood how neurons form and delete synapses over time.
For a long time, Hebbian Plasticity~\cite{hebb2005organization} was the dominant opinion.
In recent times, however, homeostatic mechanisms---in which neurons pursue a stable state and thus the whole brain reaches an equilibrium---have been suggested and shown to be accurate~\cite{butz2013,Dammasch1986}.
One of these mechanisms, the Model of Structural Plasticity~\cite{butz2013}, predicts the recovery of lesions in mice very well.
In a recent publication, Rinke et al.~\cite{Rinke2018} have reduced the quadratic complexity of solving the pairwise interactions to $O(n \cdot \log^2 n)$.
They achieved this by utilizing the Barnes--Hut algorithm~\cite{barnes1986hierarchical}, which has been developed to approximately solve pairwise interactions, which is popular in the context of physics.

We propose another approximation for the pairwise interactions based on the fast multipole method (FMM)~\cite{ROKHLIN1985187}.
While Barnes--Hut calculates point-area interactions, FMM calculates area-area interactions, reducing the complexity from quadratic to linear.
Current in vivo imaging techniques such as~\cite{chen2021} cannot precisely locate where synapses begin and end; they can only trace them to a certain area.
This, together with the fact that we do not know exactly why a particular neuron formed a (long-reaching) synapse and not its direct neighbor, gives us confidence that this approximation is reasonable. 

In this publication, we build on the Barnes--Hut approximation and utilize their distributed algorithm to implement our approximation in terms of the fast multipole method. 
To summarize, our main contributions are:
\begin{itemize}
    \item We integrated the fast multipole method into an existing parallel neuron simulation and replaced the Barnes--Hut algorithm, which was responsible for finding synapses.
    \item We reduced the theoretical complexity from $O(n/p \cdot \log^2 n)$ to $O(n/p + p)$, when $n$ is the number of input neurons and $p$ the number of MPI ranks.
    \item We measured the influence on performance in practice for different numbers of computing nodes.
\end{itemize}

The remainder of this paper is structured as follows. 
We firstly review related work in Section~\ref{sec:related} before we explain relevant background in Section~\ref{sec:background}.
Afterward, we present our algorithm in Section~\ref{sec:algorithm}, and analyze it in terms of theoretical and practical run time with multiple compute nodes in Section~\ref{sec:evaluation}.

\section{Related Work}
\label{sec:related}
There are many brain simulators freely available, for example, C2~\cite{ananthanarayanan2009}, NEST~\cite{Gewaltig:NEST}, and The Virtual Brain~\cite{sanz2013}.
They allow initial connectivity of neurons to be inserted, and during the simulation, they may strengthen and weaken those. 
However, they do not create new connections.\\
This way, their connectivity update step has complexity $O(n \cdot m)$, where $n$ is the number of neurons and $m$ is the number of synapses per neuron. 
The latter term is most often bounded (e.g., at most 1000 synapses per neuron), and thus, it is linear in the number of neurons.

Structural plasticity---the way in which neurons grow new and delete old synapses---has gained track in recent years; see~\cite{van2017rewiring} for an overview of the current state of the art.
A simple model for structural plasticity, proposed by van Ooyen et al.~\cite{VANOOYEN1994245}, achieves this by defining the outreach of a neuron to be a circle around its center.
The individual neurons form synapses proportional to the overlapping area whenever two such circles overlap.
This model requires linear time for the connectivity update, but it lacks the possibility of connecting neurons while omitting a third neuron between them.
The Model of Structural Plasticity (MSP)~\cite{butz2013}, on which this publication is based, overcomes this limitation by calculating the connection probability dependent on the distance of neurons.

In the MSP, the likelihood of a synapse forming between two neurons with positions $p_1, p_2$, is proportional to $\texttt{exp}(- ||p_1 - p_2||^2_2 / \sigma)$, with a scaling constant $\sigma > 0$.
This way, the greater the distance between two neurons is, the smaller the likelihood of a connection between them is.
Rinke et al.~\cite{Rinke2018} used this insight to approximate the influence of a whole area of neurons far away with the Barnes--Hut algorithm~\cite{barnes1986hierarchical}.
They achieved this by inserting the neurons into an octree and calculating the attraction to inner nodes whenever possible (thus skipping the need to calculate the attraction to all neurons in the induced subtree). 
This way, they reduced the complexity of calculating $O(n^2)$ interactions ($n$ again the number of neurons) to $O(n \cdot \log^2 n)$.

The fast multipole method (FMM)~\cite{greengard1991,ROKHLIN1985187} is another way of approximating pairwise interactions.
Instead of only combining the affecting elements (neurons, particles, etc.) at the calculation's source side, they also group the affected elements at the target side.
This way, they can approximate the pairwise interactions in linear time using Hermite and Taylor expansions~\cite{eylert2014praktische,friedrich2020numerische}.
This is used quite successfully in physics, including astrophysics~\cite{Dehnen.2014} and particle simulation~\cite{AMBROSIANO1988117}.
There exists many accelerated FMM implementations both with GPUs and MPI, for example~\cite{GUMEROV20088290,6375552,YOKOTA2013445}.
However, they focus on a fixed-level attraction, i.e., in contrast to us, they don't need to resolve the attractions down to a object--object level.

\section{Background}
\label{sec:background}
In this section, we repeat the arguments and definitions from previous publications to be partially self-contained.
This includes the initial publication for the Model of Structural Plasticity~\cite{butz2013}, the one that introduced the fast multipole method we use~\cite{greengard1991}, and the publication that introduced the Barnes--Hut approximation to the MSP~\cite{Rinke2018}.

\subsection{The Model of Structural Plasticity}
\label{sec:msp}
The Model of Structural Plasticity~\cite{butz2013} describes how neurons change their plasticity over time, i.e., how they form new and delete old synapses.
It consists of three different phases:\\
The update of electrical activity, the update of synaptic elements, and the update of synapses. 
An overview of the used model parameters can be seen in Table~\ref{tab:model_parameters}.

\paragraph{Update of Electrical Activity}
In the step to update the electrical activity, all neurons calculate their current activity.
This can be done by neuron models such as the one proposed by Izhikevich~\cite{izhikevich2003simple}, the FitzHugh--Nagumo model~\cite{izhikevich2006fitzhugh}, or as in our case, a Poisson spiking neuron model (the same as in~\cite{Rinke2018}).

In our model, the activity, on the one hand, strives exponentially to a resting potential (resting potential: 0.05, constant of decay: 5); on the other hand, it is constantly increased by a small background activity (0.003) and the input of all connected neurons (those neurons that form a synapse from their axon to the dendrite of the neuron in question) that spiked in the last update step by a fixed amount (5e-4).
Then a uniformly distributed value from $[0, 1]$ is drawn, and if this value is smaller than the current activity, the neuron spikes.
If a neuron spikes, it does not spike again for a fixed number of steps (refractory period: 4).

\paragraph{Update of Synaptic Elements}
In the update step, each neuron updates its intercellular calcium level.
The calcium level decays exponentially (constant of decay: 1e-5), while if a neuron spiked in that simulation step, it is increased by a fixed value (1e-3).

After the neuron has updated its calcium, it uses this to determine the amount of change to its synaptic elements.
We use the same Gaussian growth curve as originally proposed in~\cite{butz2013}, setting the right intersection (the target value) to 0.7, the left intersection (the point at which the elements start to grow) to 0.4 for axons, and 0.1 for dendrites, and the scaling parameter to 1e-4 (maximum attained value).
The neuron updates the number of axons and dendrites by the calculated amount.


\paragraph{Update of Synapses}
Every time a neuron updates its synapses (once every 100 updates of activity and synaptic elements), it checks its number of synaptic elements.
If it now has fewer elements than synapses (the elements are continuous, the synapses discrete), it chooses synapses randomly, notifies the connected neurons, and deletes them. 
It does so for both the dendrites with the incoming synapses and the axons with the outgoing synapses.

After the deletion phase, if a neuron has at least one vacant axon, it searches for another neuron with one vacant dendrite to connect to.
For each vacant axon ($i$), it calculates the probability of connecting to a vacant dendrite ($j$) by 
\begin{align}
     K(i, j) = \texttt{exp} \Big(  \frac{-||\texttt{pos}_i - \texttt{pos}_j||^2_2}{\sigma} \Big)
     \label{eq:msp:attraction}
\end{align}
and chooses one such vacant dendrite ($\sigma = 750$ as in~\cite{butz2013}).
These requests are gathered (this takes quadratic time) and sent to the neurons with vacant dendrites.
Those resolve potential conflicts, and a new synapse is formed whenever possible.

\begin{table}
    \centering
    \begin{tabular}[htb]{|l|c|c|}
        \hline
        Name & Symbol & Value \\
        \hline
        Resting potential & $x_0$ & 0.05 \\
        Membrane potential constant of decay & $\tau_x$ & 5 \\
        Background activity & $I$ & 0.003\\
        Increase in calcium per spike & $\beta$ & 5e-4 \\
        Calcium constant of decay & $\tau_{\text{Ca}}$ & 1e-5 \\
        Gaussian growth curve right intersection & $\epsilon$ & 0.7 \\
        Gaussian growth curve left intersection (axons) & $\eta_{A}$ & 0.4 \\
        Gaussian growth curve left intersection (dendrites) & $\eta_{D}$ & 0.1 \\
        Growth scaling parameter &$\mu$ & 1e-4 \\
        Probability kernel standard deviation & $\sigma$ & 750 \\
        \hline
    \end{tabular}
    \caption{Overview of the model parameters used in the executions and tests. A more detailed description of the model and the parameters can be found in~\cite{butz2013}.}
    \label{tab:model_parameters}
\end{table}
\newpage

\subsection{A Distributed Octree}
In~\cite{Rinke2018}, Rinke et al. introduced a distributed octree to overcome the memory limitations inherent to large simulations. 
The problem is that only a limited number of octree nodes can be held in memory. 
In order to be able to simulate more neurons and thus achieve the desired order of magnitude, many MPI ranks are required. 
They recursively divide the simulation domain into eight cells until a cell contains at most one neuron.
Inner nodes of the octree store the sum of vacant elements of all their children and the combined position (the centroid), which is just the weighted average position of the children.
The octree is updated in a step-wise fashion:
All ranks update their subtrees, then exchange the branch nodes, and calculate the shared upper portion afterward.
They insert all neurons into a spatial octree, where every MPI rank is responsible for 1, 2, or 4 subtrees.
All ranks share the same upper portion of the octree (heights 0 to $\log(8, p)$ where $p$ is the number of ranks), and if a rank $i$ requires information of a neuron on rank $j$, it downloads them lazily.

\subsection{Mathematical Formulation of the Fast Multipole Method}
\label{math}
Assuming there is a set of points in space, we consider a split into $M$ sources $s_1, \dots, s_M$ (in our case the neurons that have a vacant dendrite) and $N$ targets (the neurons that have a vacant axon) $t_1, \dots, t_N$.
If a neuron has vacant axons as well as vacant dendrites, it is included in both sets $M$ and $N$.
The general form of an n-body problem is~\cite{Beatson1997}:
\begin{align}
    	u(t) = \sum_{i=1}^M \omega_i \cdot K(t, s_i) \label{eq1}
\end{align}

\begin{center}
	$\omega_i \in \mathbb{R}$: The weight of the point $s_i$.\\
	$K: (\mathbb{R}^3 \times \mathbb{R}^3) \rightarrow \mathbb{R}$: A kernel that calculates the interaction between $t$ and $s_i$.\\
\end{center}

This formula gives the total attraction $u(t)$ for a vacant axon.
If a neuron has more than one vacant axon, we multiply $u(t)$ with that number.
In order to calculate $u(\cdot)$ for every target $t_j$, this function must be calculated $N$ times, which results in a total complexity of $O(N \cdot M)$. If every source is also a target, i.e., $N=M$, this scales quadratically.

The general form of Equation~\ref{eq1} fits the attraction formula of the MSP (cf. Equation~\ref{eq:msp:attraction}), with $\omega_i$ being the number of vacant dendrites of a neuron and $u(t)$ being the force of attraction to an axon of a neuron at position $t$.

\paragraph{Notation}
A multi-index $\alpha = (n_1, n_2, n_3)$ is a tuple of three natural numbers (including zero).
For any multi-index and any vector
$t = (x, y, z) \in \mathbb{R}^3$, we define the following operations:
\begin{align}
|\alpha| &= n_1 + n_2 + n_3 \\
\alpha! &= n_1 ! \cdot n_2 ! \cdot n_3 ! \\
t^{\alpha} &= x^{n_1} \cdot y^{n_2} \cdot z^{n_3}
\end{align}
For our adapted fast multipole method we often use multi-indices in combination with sums. 
For example, $\sum_{\alpha \geq p}$ or $\sum_{0 \leq \alpha \leq p}$ stands for three nested sums with $n_1, n_2, n_3 \geq p$ or $0 \leq n_1, n_2, n_3 \leq p$, respectively.


\paragraph{Approximations of Attraction Kernel}
In general the MSP sums over Gaussian functions.
We can approximate the attraction of multiple neurons in a box and group the sources and targets together.
For each such box (S for a box of sources, T for a box of targets), we need to calculate the centroid with respect to its sources $s_C$ and its targets $t_C$.
Using the function $h(\alpha, x)$ (the same as in~\cite{greengard1991} Equation~8) and $\delta=\sigma^2$, Equation~\ref{eq1} with the Gaussian kernel of Equation~\ref{eq:msp:attraction} has the following Taylor series (using a multi-index $\beta$):
\begin{equation}
\begin{split}
u({t}) 
    &= \sum_{0 \leq {\beta}} B_{{\beta}} \cdot \Biggl(\frac{{t} - {t_C}}{\sqrt{\delta}}\Biggr)^{{\beta}} \\
B_{{\beta}} &= \frac{(-1)^{|{\beta}|}}{{\beta} !} \cdot \sum_{j=1}^M \omega_j \cdot
	h\Biggl({\beta},\frac{{s_j} - {t_C}}{\sqrt{\delta}} \Biggr) 
\label{eq:taylor}
\end{split}
\end{equation}
%
%

We can truncate the outer series from Equation~\ref{eq:taylor}, i.e., sum only up to some fixed $\beta$.
The approximation error depends on the box side length, which is determined by the box's level in the octree. 
Furthermore, the number of calculated terms and the number of sources that are approximated have an influence on the approximation error.
Overall, the calculation of one coefficient $B_\beta$ from Equation~\ref{eq:taylor} has complexity $O(M)$, and crucially they are shared for all targets in the box. 
In addition, Equation~\ref{eq:taylor} must be calculated for $N$ target points with $k$ coefficients. 
This results in a complexity of $O(k \cdot M + k \cdot N)$ for a interaction between one target box and one source box.

Alternatively, we can also approximate Equation~\ref{eq:msp:attraction} with Hermite coefficients $A_\alpha$, where the same argument as before applies (with multi-index $\alpha$).
The complexity to calculate this expansion is also $O(k \cdot M + k \cdot N)$ for $k$ coefficients:
\begin{equation}
\begin{split}
    u({t}) &= \sum_{0 \leq {\alpha}} A_{{\alpha}} \cdot h \Biggl( {\alpha}, \frac{{t}- {s_C}}{\sqrt{\delta}}\Biggr) \\
A_{{\alpha}} &= \frac{1}{{\alpha} !} \cdot \sum_{j=1}^M \omega_j \cdot \Biggl( \frac{{s_j} - {s_C}}{\sqrt{\delta}} \Biggr) ^{{\alpha}} 
\label{eq:hermite}
\end{split}
\end{equation}
%
%

\section{Algorithm Description}
\label{sec:algorithm}
To determine which neurons form synapses with each other, we must calculate the forces of attraction between the target and source neurons (note here that a ``source'' and ``target'' are used differently in the literature: The ``source'' neuron is the one with the axon, however, it is the ``target'' of the attraction). 
Therefore, we create an n-body problem on top of the kernel in Equation~\ref{eq:msp:attraction} in order to apply the series expansions already presented:
\begin{align}
u(t) = \sum_{j=1}^M \omega_j \cdot \exp \left( \frac{-||t-s_j||_2^2 }{\sigma^2} \right), \label{directGauss}
\end{align}
where $\omega_j$ is the number of vacant dendritic elements of the $j$-th neuron. 
Furthermore, we use the same distributed octree as in~\cite{Rinke2018}. 
In our version---compared to the Barnes--Hut inspired one---we also need to calculate the centroid of the inner nodes with respect to the axons.
For this, we increased the size of the octree nodes from 200 Bytes to 264 Bytes (2x 32 Byte for the axon positions, which consist of 24 Bytes vector and a flag).

\begin{algorithm} 
\lstset{basicstyle=\ttfamily\footnotesize,breaklines=true}
\begin{lstlisting}[numbers=left, stepnumber=1, language= pseudo_lan]
find_synapses():
    init(stack)

    while (! stack.empty()):
        source_node, target_node = stack.pop()

        if (source_node.is_leaf && target_node.is_leaf):
            form_synapse(source_node, target_node)

        else if (source_node.is_leaf):
            new_target = choose_target(source_node, target_node.children)
            stack.push(source_node, new_target)

        else if (target_node.is_leaf):
            new_source = choose_source(source_node.children, target_node)
            stack.push(new_source, target_node)
        
        else: foreach (new_source in source_node.children)
                new_target = choose_target
                    (new_source, target_node.children)
                stack.push(new_source, new_target)
\end{lstlisting}
\caption{Pseudo code of the method \texttt{find\_synapses}. \texttt{choose\_target} and \texttt{choose\_source} are shown in Algorithm~\ref{alg:choose}.}
\label{alg:synapses} 
\end{algorithm} 

Algorithm~\ref{alg:synapses} shows the implementation for finding suitable neurons. 
For the initialization of the stack (Line~2), we first collect all roots of the subtrees and then find another subtree-root as the target for each of them, as described in the paragraph below. 
We push the source--target pairs onto the stack and then process the elements of the stack until it is empty.
Whenever we want to form a synapse, we save the source and target ids and send them to the MPI rank of the target.
Each rank collects these requests, chooses locally which to accept (to avoid too many synapses, e.g., five axons want to connect to two dendrites), and sends the answers back.

\paragraph{Initialization of the Stack}
Every MPI rank does the following:
For each of its subtree-roots, choose a target with \texttt{choose\_target} (cf. Algorithm~\ref{alg:choose}) with the global root as initial target.
Fix the subtree-root and only unpack the targets until the target is at the same level as the subtree-root. 
Put these pairs on the stack.

\paragraph{Choice of Target Node}
This method calculates the attractiveness of the target neurons to the source neuron.
It does so by first determining if it needs to evaluate the formula directly or if it can use an approximation (Taylor or Hermite).
It then calculates the attractiveness of the children of the target and chooses one randomly, proportional to their attractiveness.
In addition, the method needs two constants \texttt{c1} and \texttt{c2}, which determine when a Taylor or a Hermite expansion is used as it is easier to evaluate the attractions directly if the number of dendrites and axons is small. 

\subsection{Complexity}
For the complexity, it is enough to determine how often \texttt{choose\_target} is called.
We assume a balanced octree and start with the serial version.
Starting at level 0, the root must determine a target for each of its children, so it calls \texttt{choose\_target} 8 times and thus spawns 8 new pairs to consider.
For each of the newly created pairs, the same applies, they spawn (up to) 8 new tasks, and in general, processing level $k$ of the octree spawns $8^k$ tasks.
As the tree is balanced its height is $\log(8, n)$ for $n$ neurons, so it spawns $8, 64, \dots, n/8, n$ tasks, i.e., linear in the number of neurons.

\newpage
\texttt{choose\_target} itself either performs a direct pair-wise calculation (with quadratic complexity) or one of the FMM approximations (with linear complexity).
In our instance, however, \texttt{choose\_target} has a constant complexity because the number of sources and targets is at most 8.

For the parallel version, we have to initialize the stack first.
This requires choosing a target node for each subtree-root on the same level, which is $\log(p)$ for $p$ the number of MPI ranks.
Once we have found the pairs, we can apply the serial version to $n/p$ neurons, so the overall complexity is $O(n/p + \log p)$.
Gathering the branch nodes beforehand has complexity $O(p)$, so the overall complexity of the connectivity update is $O(n/p + p)$.

\begin{algorithm}[h]
\lstset{basicstyle=\ttfamily\footnotesize,breaklines=true}
\begin{lstlisting}[numbers=left, stepnumber=1, language= pseudo_lan]
choose_target(source_node, target_node):
    probabilities = [ ]

    for (i = 0; i < target_node.number_children; i++):
        child = target_node.children[ i ]

        if (source.is_leaf || child.is_leaf):
            probabilities[ i ] = direct_calculation()

        else if (child.get_number_dendrites() > c1 
           && source_node.get_number_axons() > c2):
            probabilities[ i ] = calculate_hermite_expanison()

        else if (child.get_number_dendrites() > c1):
            probabilities[ i ] = calculate_taylor_expanison()

        else: 
            probabilities[ i ] = direct_calculation()
    
    total_probability = sum(probabilities)
    rand = uniform_random(0, total_probability)
    index = upper_bound(probabilities, rand)

    return target_node.children[index]
\end{lstlisting}
\caption{\texttt{choose\_target} calculates the probability for \texttt{source\_node} to connect to each child of \texttt{target\_node}. It chooses the method based on the number of vacant axons and dendrites.
It then picks one target neuron randomly with chances proportional to the calculated probabilities.
\texttt{choose\_source} works analogously with swapped roles of \texttt{source\_node} and \texttt{target\_node}.
In our case, \texttt{c1 = 70} and \texttt{c2 = 70}.}
\label{alg:choose} 
\end{algorithm}

\FloatBarrier
\section{Evaluation}
\label{sec:evaluation}
Our proposed algorithm trades time against freedom of choice; the complexity of finding the target neurons is lower than for the adapted Barnes--Hut algorithm, however, we lose some freedom of choice for the synapses.
Previously, with the Barnes--Hut algorithm, each axon searched its own dendrite, so if a neuron had two vacant axons, they could connect to dendrites with a large distance between them as they could be in different nodes.
In our algorithm, both axons are always in the same box, so their choice will be the same throughout the whole process.
This also affects axons on neurons that are close to each other---if they are in the same box on level $l$, their choice of boxes coincides on every level $i = 0, \dots, l-1$.
This, in turn, means that every neural network that was calculated with the fast multipole method can also be calculated with the Barnes--Hut algorithm, but not vice versa.

All calculations for this research were conducted on the Lichtenberg 2 high-performance computer of the TU Darmstadt.
One compute node has 2 Intel Xeon Platinum 9242 processors (with disabled hyper-threading), 384 GB main memory, and the interconnection is a 100 GBit/s InfiniBand.
We always tested the algorithm with 500\,000 simulation steps (5000 connectivity updates) and a network with only excitatory neurons.

\begin{figure}

    \centering
    \includegraphics[width=1\linewidth, trim=2cm 0.2cm 2cm 0.7cm]{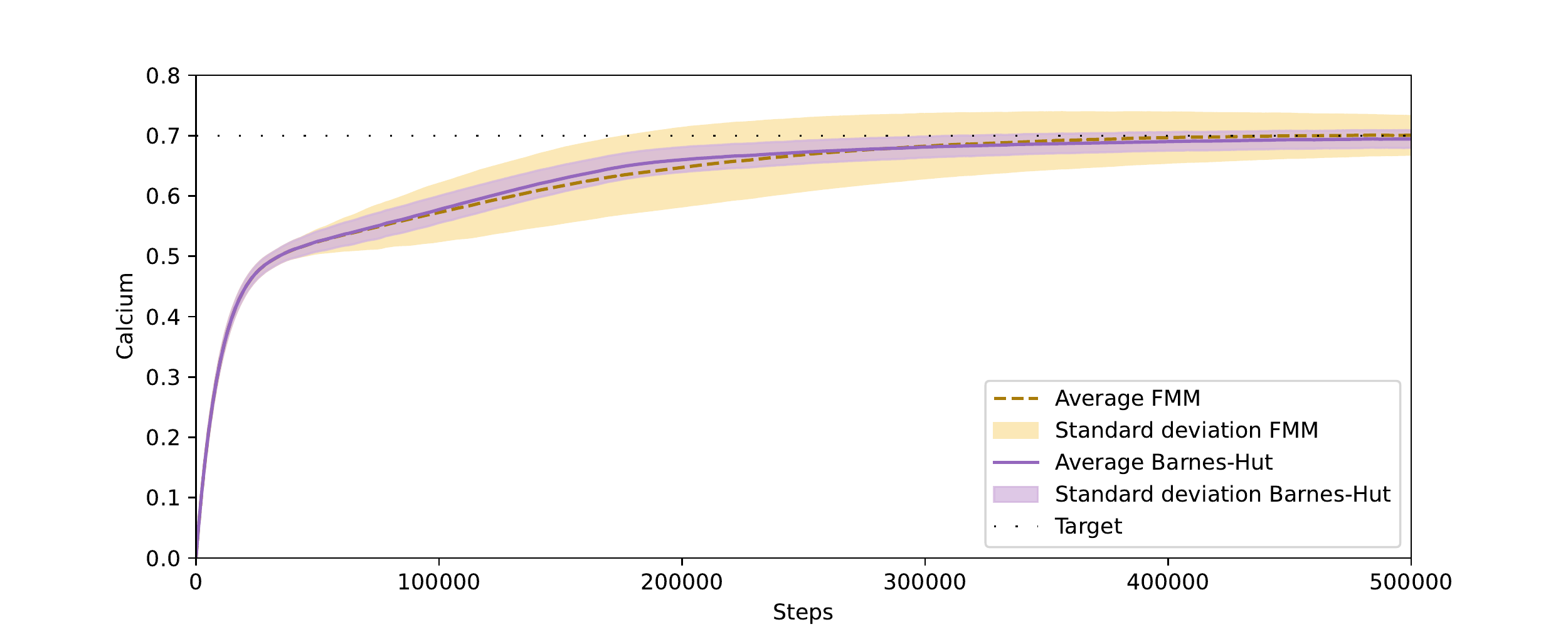}
    \caption{The average calcium (solid lines) and its standard deviation (light area) for both algorithms (Barnes--Hut in solid purple, fast multipole method in dashed orange). The target calcium is 0.7 (thinly dashed black line). $p = 64$ MPI ranks, $n = 320\,000$ neurons, and $500\,000$ simulation steps ($5000$ connectivity updates).\\}
    \label{fig:calcium}
    
    \centering
    \includegraphics[width=1\linewidth, trim=2cm 0.2cm 2cm 1cm]{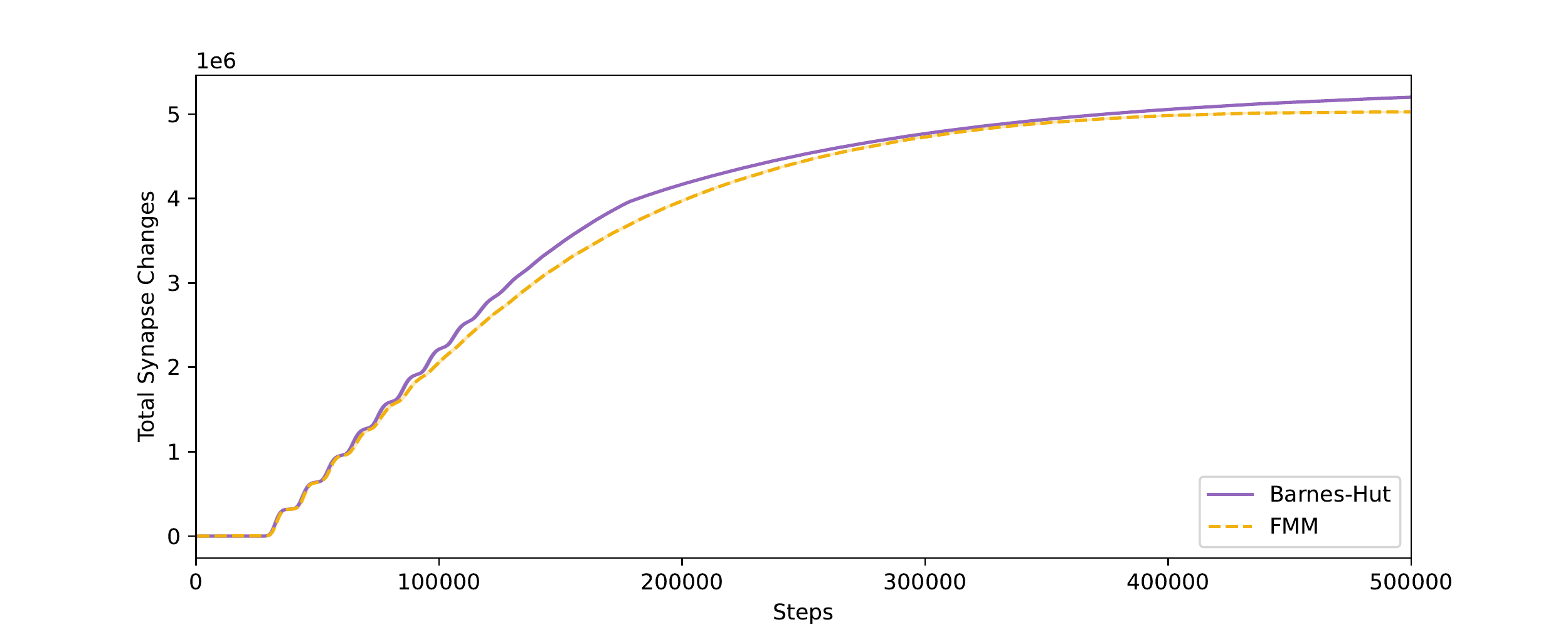}
    \caption{The total number of synapses for both algorithms. $p = 64$ MPI ranks, $n = 320\,000$ neurons, and $500\,000$ simulation steps ($5000$ connectivity updates).\\}
    \label{fig:synapses}
    
     \centering
    \includegraphics[width=\linewidth, trim=0cm 0.8cm 0cm 0.8cm]{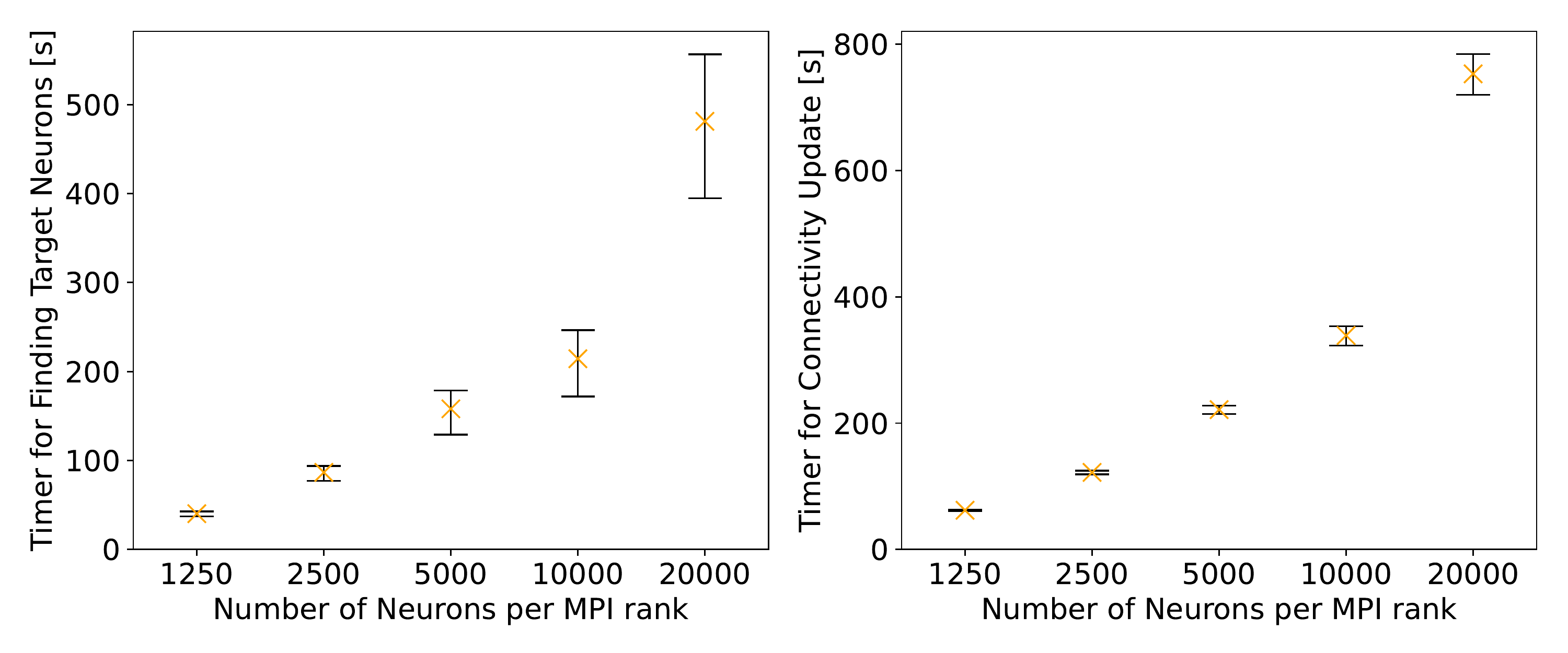}
    \caption{The timings for the strong scaling experiments with $p=64$ MPI ranks and $500\,000$ simulation steps ($5000$ connectivity updates). We give the minimum, average, and maximum time across the different ranks. All timings are in seconds.}
    \label{fig:strong_scaling_times}
\end{figure}

From a neuroscientific point of view, we investigated the following observable metrics for both the Barnes--Hut and the fast multipole method inspired algorithms:
\begin{enumerate}
    \item The average calcium concentration of the neurons (together with the standard deviation) to see how well both algorithms allow the neurons to reach a local equilibrium.
    \item The number of formed synapses to see how well the overall simulation reaches a global equilibrium.
\end{enumerate}
Figure~\ref{fig:calcium} shows the average calcium of the neurons (together with its standard deviation), and Figure~\ref{fig:synapses} shows the total number of created synapses for one run of $p = 64$ MPI ranks and $n = 320\,000$ neurons.
With our proposed algorithm, the average calcium is nearly indistinguishable from the Barnes--Hut algorithm, however, its standard deviation is slightly higher.
Our algorithm trails the previous version slightly when it comes to the total number of formed synapses.
This is due to more collisions, resulting in more rejections, so we need more simulation steps to connect all vacant axons.
Furthermore, the total number of synapses is less for our algorithm.
The reason is that a neuron generally grows more dendrites than axons.
If the synapses now cluster more (due to the restricted freedom of choice), some neurons receive more synapses than they want---so they delete some again.

\newpage
We evaluated the strong-scaling behavior of our algorithm with $p = 64$ MPI ranks and $n = 1\,250, 2\,500, 5\,000, 10\,000, 20\,000$ excitatory neurons per MPI rank.
Figure~\ref{fig:strong_scaling_times} shows the minimum, average, and maximum time for a simulation across all MPI ranks.
We have conducted these simulations five times and found that the generated timings are very stable concerning the repetitions.
When calculating the coefficient of variation (standard deviation divided by average) for these repetitions, it is consistently below 1\%.
Doubling the number of neurons per rank scales the time of the connectivity update by approximately 1.96, 1.81, 1.53, and 2.22, and the time it takes to find the targets by 2.15, 1.82, 1.35, and 2.25.
Overall, these timings suggest to us a good strong-scaling behavior.

\begin{figure}
    \centering
    \includegraphics[width=1\linewidth - 2 cm, trim=0cm 1cm 0cm 1cm]{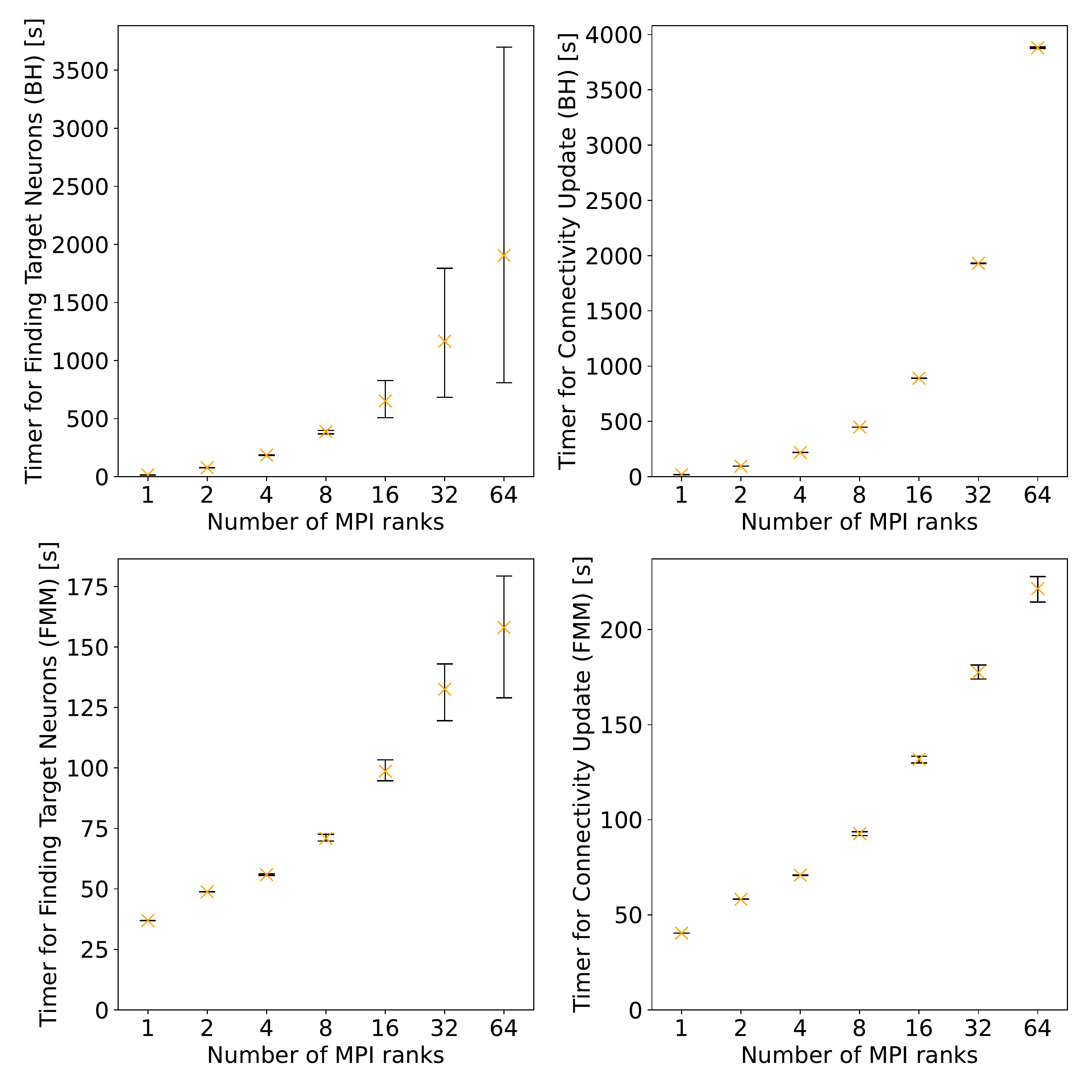}
    \caption{The timings of different methods of the simulation for $p = 1, 2, 4, 8, 16, 32, 64$ MPI ranks and $500\,000$ simulation steps ($5000$ connectivity updates).. For each method we give the minimum, average, and maximum time across the different MPI ranks. All timings are in seconds.\\}
    \label{fig:weak_scaling_times}
    
    \centering
\begin{minipage}[b]{.35\linewidth}
      \includegraphics[width=\linewidth, trim=0cm 0.5cm 0cm 0cm]{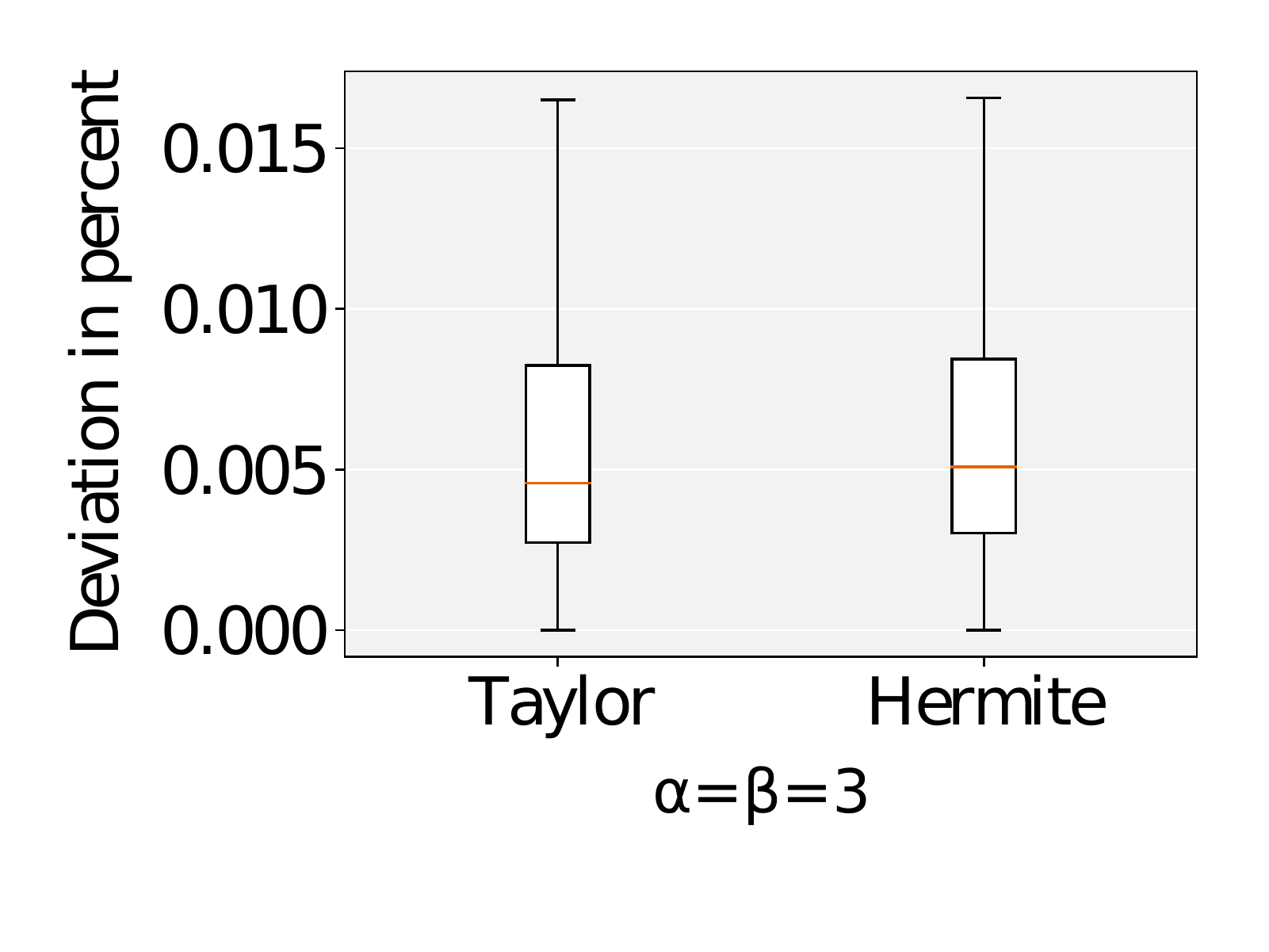}
   \end{minipage}
\begin{minipage}[b]{.35\linewidth}
      \includegraphics[width=\linewidth, trim=0cm 0.5cm 0cm 0cm]{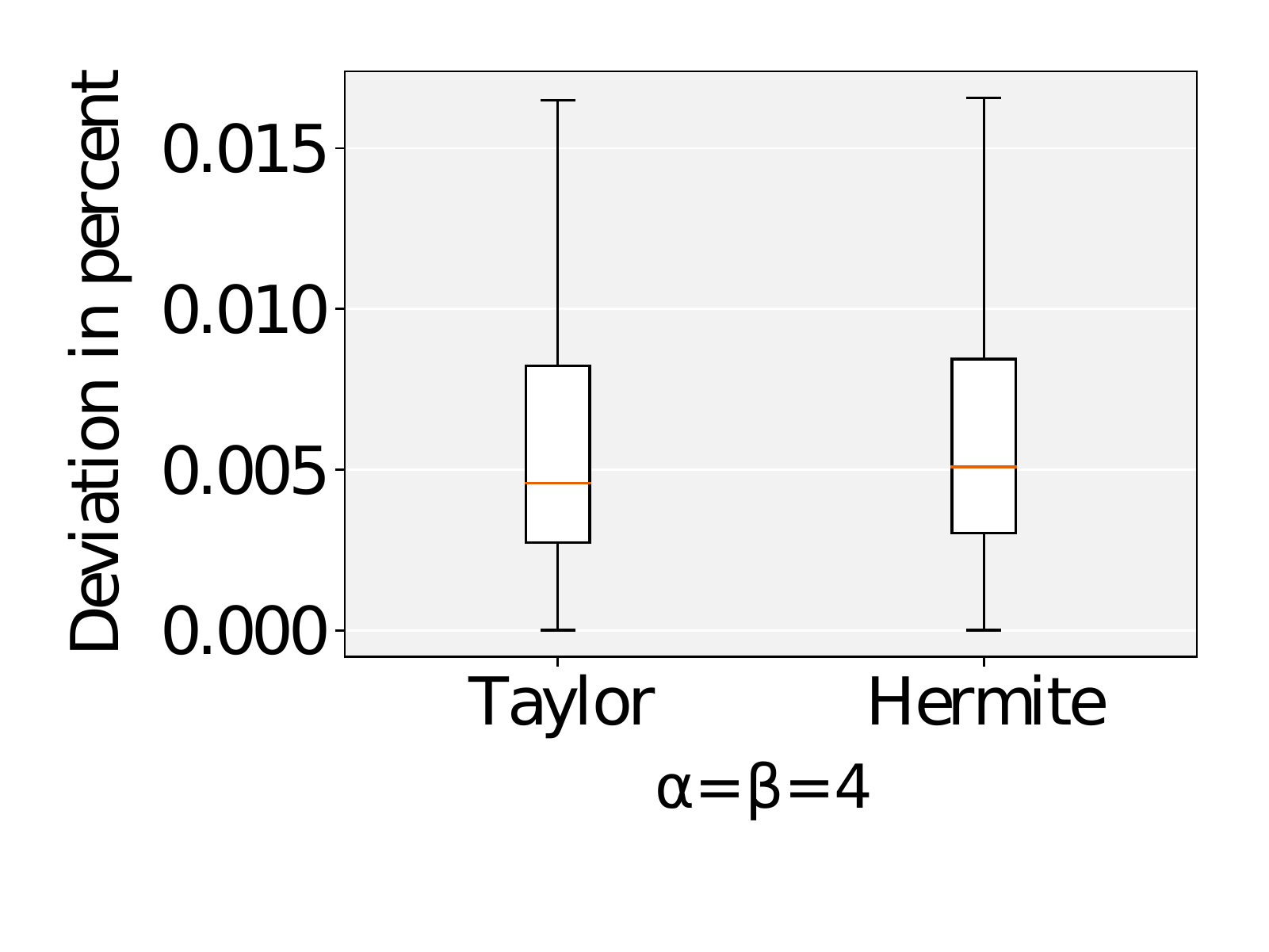}
\end{minipage}
\begin{minipage}[b]{.35\linewidth}
      \includegraphics[width=\linewidth, trim=0cm 2cm 0cm 1cm]{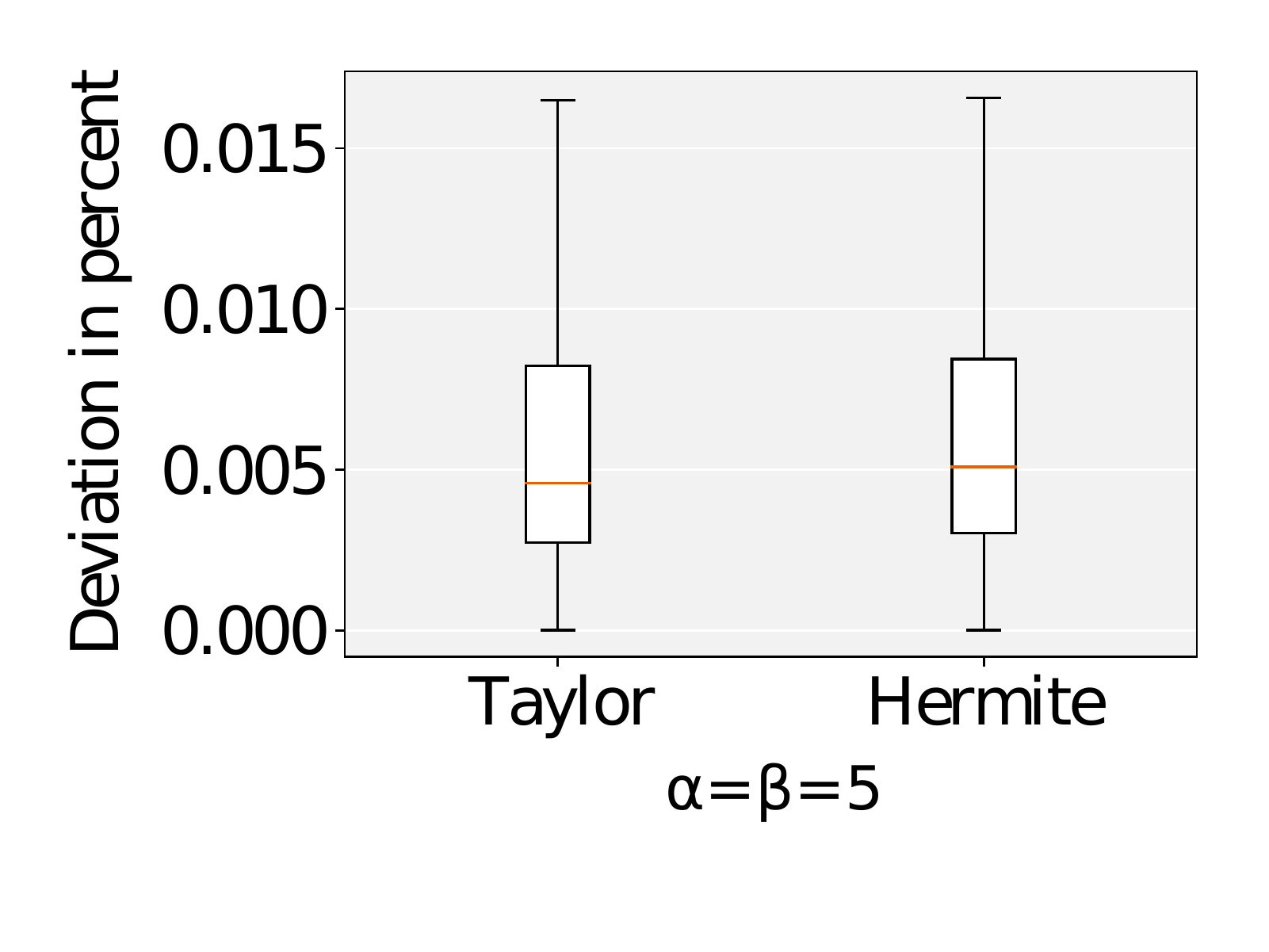}
   \end{minipage}
   \begin{minipage}[b]{.35\linewidth}
      \includegraphics[width=\linewidth, trim=0cm 2cm 0cm 1cm]{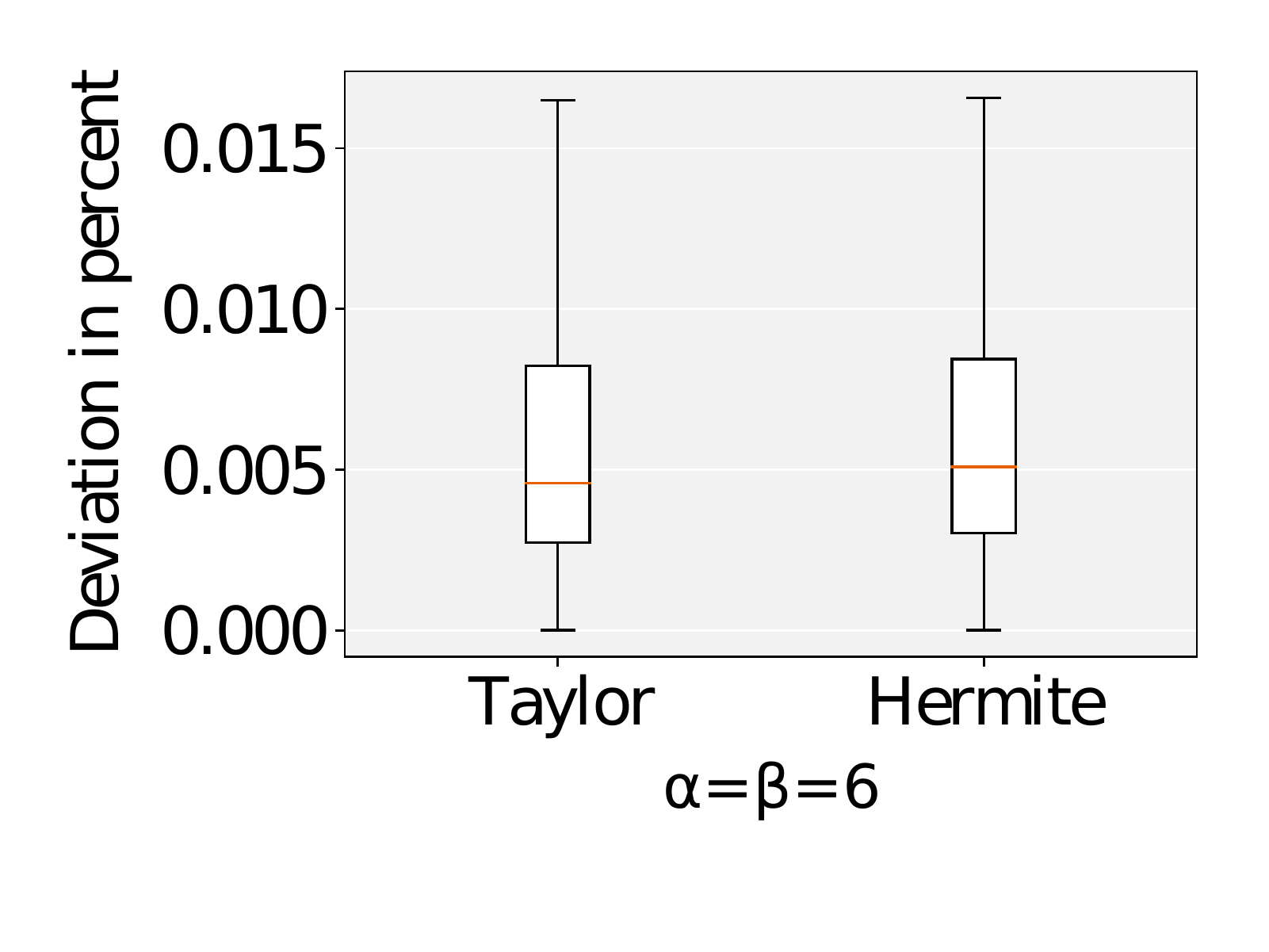}
\end{minipage}
\caption{The deviation in percent between the directly evaluated attraction and the corresponding Hermite and Taylor expansions, gathered from $12\,188$ representative boxes.
The red line is the median, the box indicates the 0.25 and 0.75 quartile, the interval indicates the minimum and maximum after removing all outliers. 
A value is an outlier if it is larger then the 0.75 quartile + 1.5 times the inter-quartile range.
The number of outliers for the Taylor expansions were 1830, 1834, 1833, and 1833, and there were consistently 1753 outliers for the Hermite expansions.
The largest outliers were below 0.125 \%.}
\label{fig:accuracy}
\end{figure}

Furthermore, we tested the weak-scaling behavior of our algorithm with $n = 5000$ neurons per MPI rank (as in the previous publication which introduced the Barnes--Hut approximation~\cite{Rinke2018}) and with $p = 1, 2, 4, 8, 16, 32, 64$ MPI ranks.
For the timings, we investigated the overall time for the connectivity update, the time it takes to find the target neurons, and for the fast multipole method also the time it takes to compute the expansions.
Figure~\ref{fig:weak_scaling_times} shows the minimum, average, and maximum time for a simulation across all MPI ranks.
We have repeated each measurement five times with no significant difference. This means that the coefficient of variation remained below 1\% in this experiment as well.
Between one tenth and one third of the time the fast multipole method spends finding the synapses is spent in the Taylor expansion; the Hermite expansion is rarely used.
The difference between the MPI ranks is low in our algorithm compared to the Barnes--Hut algorithm.
Besides network communication noise, this difference is caused by neurons choosing partners close to or far away from others.
Per connectivity update, we cache the already fetched octree nodes from other MPI ranks.
This way, our algorithm profits from the locality of target choices compared to the Barnes--Hut algorithm.
Overall, the new connectivity update is significantly faster, and the scaling behavior fits the broad expectation of $O(n/p + p)$.

Lastly, we evaluated the influence of the parameters $\beta$ from Equation~\ref{eq:taylor} and $\alpha$ from Equation~\ref{eq:hermite}, i.e., the points at which we cut of the evaluation of the infinite series.
For this, we have conducted $12\,188$ representative calculations for each expansion, as well as the direct evaluation.
Figure~\ref{fig:accuracy} displays the results, showing that our cut-off point with $\alpha = \beta = (3, 3, 3)$ is well chosen and more terms do not enhance the accuracy significantly.

\FloatBarrier

\section{Conclusion}
This work aimed to replace the Barnes--Hut algorithm in an existing neuron simulation with multiple computing nodes with the fast multipole method of lower complexity and to measure the influence on performance. 
We achieved a theoretical complexity of $O(n/p + p)$, when $n$ is the number of input neurons and $p$ the number of MPI ranks, which is lower than the previous complexity of $O(n/p \cdot \log^2 n)$. 
In addition, the algorithm presented here is faster in practice on multiple computing nodes, exhibits a good strong-scaling behavior, and additionally, the rank-to-rank variation shrank significantly.
Also, the internal calcium concentration and the formation of synapses behave very closely to the original simulation. 
However, there are aspects in which the algorithm presented here is inferior to the Barnes--Hut algorithm. 
The storage space consumption has increased by 32 \%, and the choices of neighboring neurons are now more similar than before. 
In addition, our algorithm needs more simulation steps to connect all vacant elements through synapses due to more collisions.

In the future, we seek to combine the variable precision of the Barnes--Hut algorithm with our proposed one, which might let neurons form connections more independently than their neighbors.
Furthermore, we plan to analyze the resulting networks with respect to the graph-topological metrics so we can assess the functionality of the networks.

\section*{Acknowledgments} 
We acknowledge the support of the European Commission and the German Federal Ministry of Education and Research (BMBF) under the EuroHPC Programme DEEP-SEA (955606, BMBF Funding No. 16HPC015). The EuroHPC Joint Undertaking (JU) receives support from the European Union's Horizon 2020 research and innovation programme and GER, FRA, ESP, GRC, BEL, SWE, UK, CHE. This research was also supported by the EBRAINS research infrastructure, funded by the European Union’s Horizon 2020 Framework Programme for Research and Innovation under the Specific GA No. 945539 (Human Brain Project SGA3), and is partly funded by the Federal Ministry of Education and Research (BMBF) and the state of Hesse as part of the NHR Program.
The authors gratefully acknowledge having conducted a part of this study on the Lichtenberg high-performance computer of TU Darmstadt.

\bibliographystyle{splncs04}
\bibliography{bibliography}

\end{document}